\documentclass[aps,pre,twocolumn,groupedaddress,showpacs,floatfix]{revtex4}
\usepackage{amsmath}
\usepackage{amssymb}
\usepackage{graphicx}

\begin{document}

\title{A self-referred approach to lacunarity}

\author{Erbe P. Rodrigues, Marconi S. Barbosa and Luciano da F.
Costa}

\affiliation{Institute of Physics of S\~ao Carlos.  University of S\~
ao Paulo, S\~{a}o Carlos, SP, PO Box 369, 13560-970, phone +55 162 73
9858,FAX +55 162 71 3616, Brazil, luciano@if.sc.usp.br}

\date{5th May 2004}

\begin{abstract} 

This letter describes an approach to lacunarity which adopts the
pattern under analysis as the reference for the sliding window
procedure.  The superiority of such a scheme with respect to more
traditional methodologies, especially when dealing with finite-size
objects, is established and illustrated through applications to DLA
pattern characterization.  It is also shown that, given the enhanced
accuracy and sensitivity of this scheme, the shape of the window
becomes an important parameter, with advantage for circular windows.

\end{abstract}


\maketitle

Several interesting natural and abstract phenomena and structures are
characterized by intricate geometries whose properties can vary
along space and/or
time~\cite{Plotnick:1996,Smith:1996,Costa:2004,Einstein:1998,Witten:1981}.
Chaotic dynamics, for instance, is known to be organized in terms of
fractal attractors~\cite{Grebogi:1987}, which are characterized by
self-similarity or self-affinity over spatial scales.  Given that
great part of the systems exhibiting particularly interesting behavior
involves such complex geometrical organizations, it becomes important
to have proper and effective measurements allowing the objective and
meaningful quantification of specific geometrical features, such as
regularity, density, self-similarity and translational invariance.
One important point to be highlighted at the outset is the fact that
such measurements are almost invariably \emph{incomplete} or
\emph{degenerated}, as a consequence of the mapping from a higher
dimensional space, where the structures `live', into a lower
dimensional space.  Therefore, while it is often unfeasible to
incorporate all information into geometrical measurements, they 
must be capable of expressing the features of particular interest 
with respect to each specific application.  For instance, the 
characterization of the distribution of empty space of different 
sizes and shapes is a major factor to be considered while
specifying the mechanical properties of a metal bar for example,
establishing an intrinsic relationship between topological/geometrical
properties and physical strength that can be to some extent captured
by its porosity value in case a parsimonious description is needed.
By providing accurate and meaningful information about the specific
geometrical properties of interest, proper measurements of complex
structures allow the construction of statistical models of the
analyzed objects and the identification of prototypes, as well as the
taxonomic organization of several types of patterns.  Such
possibilities are important not only for practical applications, but
also for theoretical studies aimed at investigating critical phenomena
and universality~\cite{Gefen:1983}.

One of the best known measurements of complex structures is the
\emph{fractal dimension}, introduced by B. Mandelbrot,
see~\cite{Mandelbrot:1983}.  Although several alternative definitions
of such a measurement have been available for a long time
(e.g.~\cite{Falconer:2003,Peitgen:2004}), they all assume self-similar
(or self-affine) symmetries while sharing the ability to quantify the
spatial `complexity' of given patterns.  Although powerful and widely
used, the fractal dimension is inherently a degenerated feature,
implying an infinite amount of distinct fractals to be mapped into the
same fractal dimension.  The concept of \emph{lacunarity}
\cite{Mandelbrot:1983,Allain:1991,Gefen:1983} has been introduced and
used as a means to complement the quantification of complex geometries
provided by the fractal dimension.  In particular, the lacunarity
quantifies the degree of \emph{translational invariance} of the
analyzed objects, with low values of lacunarity indicating high levels
of such an invariance.  A particularly representative illustration of
the potential of the combined use of the fractal dimension and
lacunarity is related to the characterization of DLA structures which,
by being organized around the initial `seed', tend to exhibit distinct
geometrical properties around that seed and also at the DLA
boundaries~\cite{Mandelbrot:1995}.

Despite the promising potential of the lacunarity as a measurement of
complex patterns, some remaining intrinsic difficulties have conspired
to impinge some degree of arbitrariness, constraining its
applications.  Of special importance is the lack of a proper procedure
to treat finite-size objects.  Indeed, while lacunarity and fractal
dimension are often considered for the characterization of
infinite/periodical structures, the treatment of finite and isolated
objects, implied by many relevant natural situations, has received
relatively little attention in the literature.

The current work investigates the use of the own analyzed pattern
as the reference for the windowing procedure underlying lacunarity
estimation.  Although such an approach has been considered
previously~\cite{Allain:1991}, the restricted conditions adopted for
its validation (Cantor dust) implied its premature dismissal.  An
interesting informal interpretation of this approach is to understand
the structure of interest as being measured by an inhabitant of the
object who, therefore, can only sample a circular region around each
of its positions.  We show in the following that this self-reference
windowing system does allow a series of superior features, including
enhanced objectivity, accuracy and sensitivity, also implying the
shape of the sliding window to become critical for proper operation.
It is shown that such an alternative procedure allows the additional
bonus of enhanced computational speed.

\begin{figure}[ht]
\begin{center}
\includegraphics[scale=0.3,angle=-90]{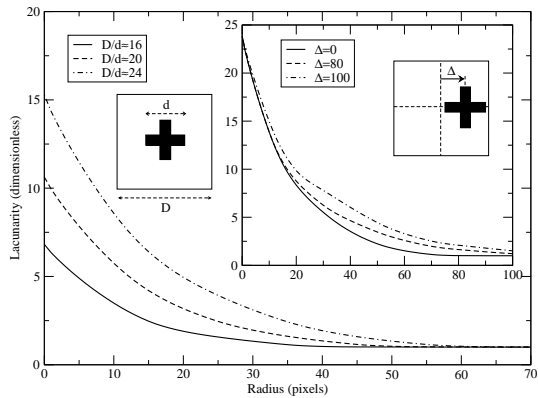}
\caption{Some aspects constraining the practical use of the standard
implementation of the lacunarity concept for finite objects
characterization.~\label{fig:drawnbacks}}
\end{center}
\end{figure}

\begin{figure}[ht]
\begin{center}
\includegraphics[scale=0.35,angle=0]{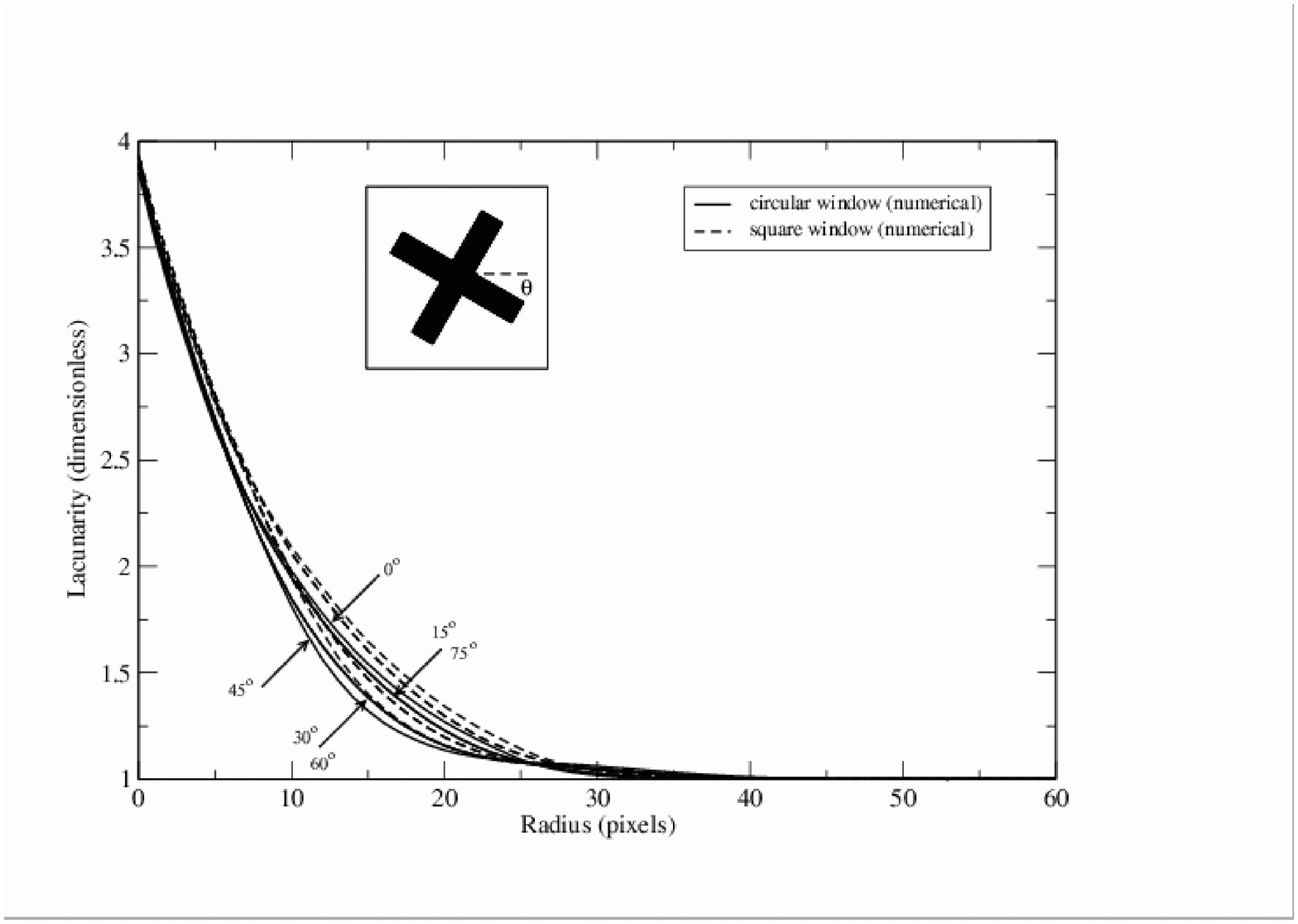}
\caption{Inconsistency under object rotation of the usual procedure for
  measuring the lacunarity concept.~\label{fig:rotavelha}}
\end{center}
\end{figure}

\begin{figure}[ht]
\begin{center}
\includegraphics[scale=0.2,angle=0]{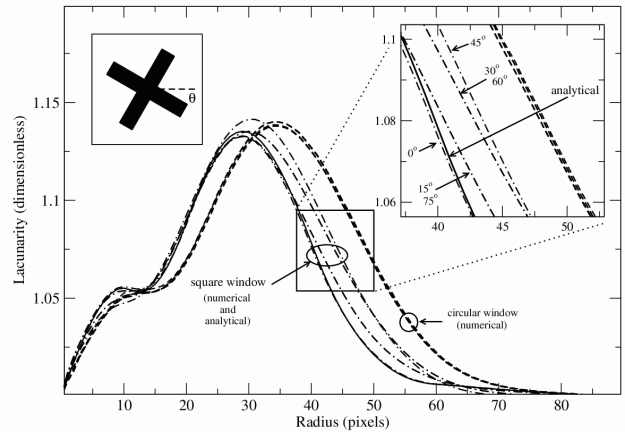}
\caption{Stability of the proposed lacunarity measurement under 
 rotation of the object under analysis.~\label{fig:stability}}
\end{center}
\end{figure}

\section{Methodology}

One of the most traditional approaches to estimate the lacunarity of a
set of objects is known as the gliding-box algorithm,
(e.g.~\cite{Allain:1991}). Provided the set under analysis is mapped
into a square lattice of dimension $L \times L$, henceforth called the
\emph{workspace}, a window of side $l$ is made to slide through the
entire lattice while the number of pixels which falls inside it is
determined and recorded.  Let $n(s,l)$ be the number of such boxes
which contain $s$ pixels and $N(l)$ be the total number of boxes of
size $l$.  The probability of finding a box of size $l$ with $s$
pixels is given by $Q(s,l)=n(s,l)/N(l)$, and the lacunarity
$\Lambda(l)$ of such a pixel distribution can be simply expressed as

\begin{equation}
\Lambda(l)=\frac{\Sigma s^2 Q(s,l)}{[ \Sigma sQ(s,l) ]^2}=\frac{\sigma_{s}^2(l)}{\overline{s}^2(l)} + 1
\end{equation}

Although popular, such a procedure involves some arbitrariness related
to the difficulty to choose the several involved parameters such as
the position and size of the workspace and the shape of the sliding
window.  Figure~\ref{fig:drawnbacks} illustrates the results of
applying the sliding approach considering a square window to the
inlaid cross.  The outset presents three lacunarity signatures
obtained for three distinct ratios between the working space and
object sizes.  It is clear from these curves that the choice of
proportionality ratio can have great effect in defining the lacunarity
values.  The inset curves were obtained for a fixed working space
size, but with the object (a cross) placed at different relative
positions.  A strong variation of the obtained lacunarity values was
again observed, indicating arbitrariness also regarding the object
position.  Therefore, the large variations implied by the above
arbitrary choices undermine the potential of the lacunarity as a
sensitive measurement of the spatial distribution of the analyzed
finite structures.  The orientation of the object under analysis
represents an additional arbitrary aspect of the traditional
sliding-window approach.

The arbitrariness identified above can be completely removed by the
use of the structure under analysis as the reference for placing the
sliding windows.  In other words, the window is placed at each of the
points of that structure, eliminating the influence of the workspace,
which can now be objectively defined by the maximum sliding-window
size and the structure under analysis.  The remaining parameters are
therefore reduced to the shape of the sliding-window and the
spatial-scale interval of the analysis (i.e. the range of window
sizes), accounting for enhanced objectiveness of the whole approach.

Figures~\ref{fig:rotavelha} and~\ref{fig:stability} illustrate the
stability of the traditional and proposed lacunarity functionals
regarding rotation. Figure~\ref{fig:rotavelha} shows the traditional
lacunarity curves obtained for rotations of the considered pattern
considering square and circular windows.  A substantial variation is
observed for both types of windows.  Figure~\ref{fig:stability} gives
the self-referred lacunarity of the object shown on the upper left for
diverse rotation angles, two sliding window shapes, as well as the
respective analytical result. In the main graph we can easily spot two
main groups: one associated with the square sliding window and
another, more tightly grouped, associated with the circular
window. Among the more widespread group, one can see a solid line
representing the analytical calculation expected for the
self-reference method, which matches precisely the numerically
evaluated curves. The inset provides a zoomed view of the variance
implied by the use of a square sliding window, which is particularly
critical if rotational invariance is required.

\section{Results}

\begin{figure}[ht]
\begin{center}
\includegraphics[scale=0.65,angle=-90]{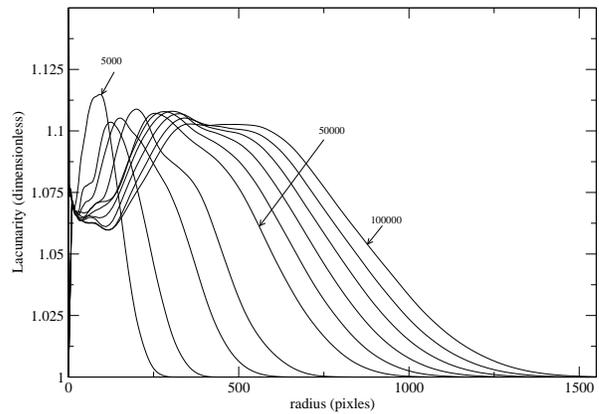}

\caption{The effect of the number of aggregated particles on the
 self-referred lacunarity value. As the number of particles grows,
 the lacunarity maximum tends to a standstill, suggesting itself as a
 prominent feature for morphological characterization of such 
 complex shapes, as exemplified by
 Figure~\ref{fig:scatter}.~\label{fig:converge}}

\end{center}
\end{figure}

An important feature of many quasi self-similar shapes is the
existence of a descriptor, such as the fractal dimension, which can
provide a characteristic signature for the shape regardless of the
number of aggregated particles. The lacunarity represents one such a
descriptor which has been proposed in order to complement the fractal
characterization and exhibits a `convergent behavior' as the number of
particles reaches a critical value~\cite{Smith:1996}.
Figure~\ref{fig:converge} illustrates such a property with respect to
the standard procedure for DLA generation~\cite{Tolman:1989}. The
behavior of the maximum of such curves suggests itself as a possible
measurement for characterizing the whole sequence of produced
individual shapes.

\begin{figure}[ht]
\begin{center}
\includegraphics[scale=0.4,angle=0]{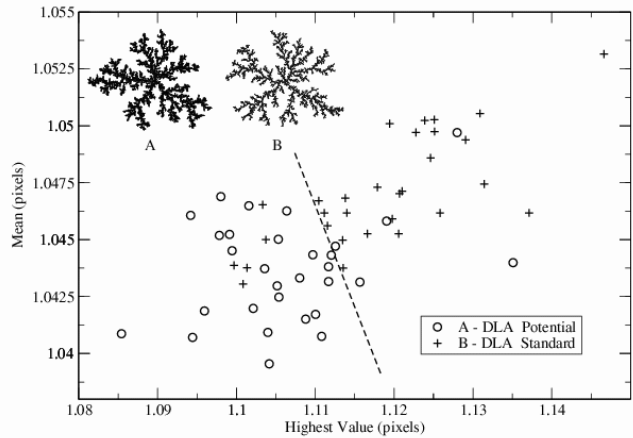}

\caption{A scatter plot defined by functionals extracted from the
 lacunarity curve, namely the global standard deviation and the local
 maximum value. The plot shows some amount of similarity among the two
 differently grown types of DLAs.~\label{fig:scatter}}

\end{center}
\end{figure}

In order to investigate the potential of the self-referred lacunarity approach
for pattern discrimination, an experiment has been carried out in which two
differently grown sets of DLA structures~\cite{Tolman:1989,Caserta:1990} with 30 samples
each, are analyzed by the self-referred approach described in this paper. The
results of this experiment are summarized in Figure~\ref{fig:scatter}, which
shows two clearly separated clusters (as illustrated by the straight dashed
frontier), with some overlap at their borders.  Such an overlap is a
consequence of some degree of similarity between the two types of structures,
which was detected by the considered measurement.  It is observed that, out of
the two considered measurements, the highest lacunarity value (represented
along the $x$-axis) contributed more effectively to the separation between the
two classes of objects.

\begin{figure}[ht]
\begin{center}
\includegraphics[scale=0.6,angle=-90]{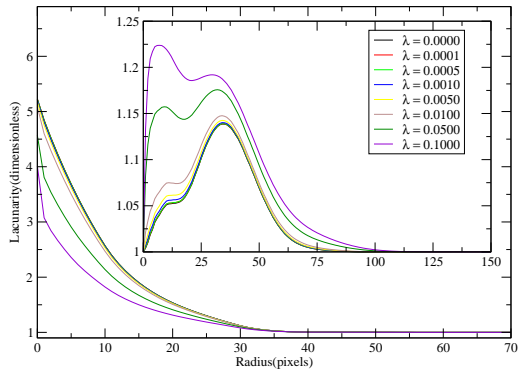}
\caption{The resilience of the proposed lacunarity against perturbation by
  Poisson noise with varying number density.~\label{fig:resilience}}
\end{center}
\end{figure}

Another important issue is related to the sensitivity of this new
approach to small variations of the object. We consider this important
perspective through an experiment where the object is perturbed by
increasing Poisson noise. The outcome of such a study is presented in
figure~\ref{fig:resilience}, which shows the self-referred lacunarity
curves obtained for several levels of noise, quantified by the
respective Poisson rates given by the respective legend.  While the
traditional lacunarity, shown in the outset graph, is characterized by
maximum relative variation of 0.73 against 0.13.  Such a result
suggests that the self-referred method present enhanced robustness
when compared to the traditional lacunarity.

\section{Comments and Conclusions}

While the traditional approach to lacunarity estimation involves
sliding a box throughout the space where the structure under analysis
is contained, the application of such a procedure implies substantial
arbitrariness when applied to general shapes and sets of objects
characterized by finite size. Of particular importance is the fact
that there is no established criterion for defining the positions of
the sliding box along the space under analysis, so that different
implementations will often converge to different results.  We have
shown that the adoption of the own objects under analysis as the
reference for positioning of the sliding window provides not only a
fully objective procedure for lacunarity estimation, but also enhances
its potential for discriminating between different classes of
patterns.  Such effects have been demonstrated with respect to the
important problem of DLA pattern formation and analysis.  In addition,
the stability of the self-referred approach has been investigated with
respect to Poisson perturbations, suggesting good
robustness. Moreover, the enhanced signature provided by the
object-referred framework considered in this article makes the choice
of the window geometry an important issue.  In particular, we have
shown that circular (spherical) windows provide superior properties
when used for self-referred lacunarity estimation by promoting the
isotropy of the analysis.  An additional advantage allowed by the
considered lacunarity definition is its substantially reduced demand
for computational resources.  As the sliding window is constrained to
the object under analysis, the total of integrations along the window
is reduced from a large area around the object to its own area, which
often imply savings of an order of magnitude.

\begin{acknowledgments}

 Luciano da F. Costa is grateful to FAPESP (process 99/12765-2), CNPq
 (308231/03-1) and the Human Frontier Science Program for financial
 support.  Marconi S. B. is thankful to FAPESP (process
 02/02504-1,03/02789-9) for financial support, and Erbe Pandini thanks
 to CNPQ for his financial support.

\end{acknowledgments}

\bibliography{lacun}

\end{document}